\documentstyle[12pt]{article}
\begin{document}
\noindent
{\bf Non-linear evolution equations with non-analytic dispersion}\\
{\bf relations in 2+1 dimensions. Bilocal approach}\\
\vspace{1cm}

\begin{tabular}{lll}
& & E.V.Doktorov\\
& &\\
& & B.I.Stepanov Institute of Physics, F.Skaryna Ave. 70 \\
& & 220072 Minsk, Republic of Belarus\\
\end{tabular}
\vspace{1cm}

\noindent
Non-linear equations in 2+1 dimensions\\

\vspace{1cm}

\noindent
\begin{tabular}{lll}
PACS & 0230 & Function theory, analysis\\
    & 0340 & Classical mechanics of continuous media:\\
    &      & general mathematical aspects\\
\end{tabular}
\newpage

\noindent
{\bf Abstract}. A method is proposed of obtaining (2+1)-dimensional
non-linear equations with non-analytic dispersion relations. Bilocal
formalism is shown to make it possible to represent these equations in a
form close to that for their counterparts in 1+1 dimensions.
\newpage
\noindent{\bf1. Introduction}\\
\medskip

\noindent
Non-linear evolution equations with non-analytic (singular) dispersion
relations (SDR equations) form an important class of equations integrable
by means of the inverse spectral transform. The Maxwell-Bloch equations
\cite{1} are the well-known representative of this class in 1+1 dimensions.
A general construction of (1+1)-dimensional SDR equations solvable via the
Zakharov-Shabat spectral problem was given by Leon \cite{2}. As was shown
by Boiti et al in an interesting paper \cite{3}, the SDR equations in 2+1
dimensions possess a number of peculiarities, the main one being the absence
of an explicit expression for the evolution linear operator $T_2=\partial_t
-W$ which enters the Lax representation. Nevertheless, this circumstance does
not prevent from a construction of soliton solutions by means of the
B\"acklund transformations. In particular, it was proposed in \cite{3} a
(2+1)-dimensional generalisation of the Maxwell-Bloch equations which had
a form of a rather complicated system of four equations. In our opinion,
such a complexity was caused by the fact that the approach realised in
\cite{3} was primarily based on the function $W$ given unexplicitly. In this
connection, it is seemed to be reasonable to propose another way of deriving
the above class of equations without making the direct use of the function
$W$.

We will consider as a primary object a spectral transform $R$ appearing in
the framework of the $\bar\partial$-method [4-7].
 Hence, the aim of
the present paper is to obtain a hierarchy of (2+1)-dimensional non-linear
equations with non-analytic dispersion relations compatible with the
linear evolution of the spectral transform $R$. We will demonstrate that
the formalism developed by Beals and Coifman \cite{8} for holomorphic
dispersion relations can be naturally adapted for equations of interest.
Our consideration relies essentially on the bilocal approach initiated by
Konopelchenko and Dubrovsky \cite{9} and elaborated to a full extent by
Fokas and Santini \cite{10,11}. It is precisely the bilocal formalism
that allows us to generate in a natural manner (2+1)-dimensional analogues
of many structures which successfully work in 1+1 dimensions. We will show
that the form of the SDR equations in 2+1 dimensions written in bilocal
variables is very close to that for equations in 1+1 dimensions. In
particular, the 'squared eigenfunction' structure typical for the
(1+1)-dimensional situation takes place also in 2+1 dimensions. In the
process of deriving a hierarchy of equations we shall not use, as distinct
from \cite{10}, an extended integral representation for the function $W$
(due to the lack of an explicit expression for it). In our view, the
proposed way of obtaining the recursion operator follows more closely the
lines of 1+1 dimensions.

In the following, we shall restrict ourselves to the consideration of the
hyperbolic spectral problem. A derivation of the relevant formulae in the
case of the elliptic spectral problem does not cause principial difficulties.
\bigskip

\noindent {\bf2. Lax representation and $\bf{\bar\partial}$-problem}
\medskip

\noindent
As a starting point in a construction of nonlinear SDR equations, we
consider a $\bar\partial$-problem on a complex plane ${\cal C}$
($\bar\partial\equiv\partial/\partial\bar k$):
\begin{eqnarray}
\bar\partial\phi(k)&=&\int\!\int\,dl\wedge
d\bar l\,\phi(l)R(k,\,l),
\qquad k,\,l\in\cal C,\\
\phi(k)&=&1+O(1/k), \quad k\to\infty, \quad \phi\in SL(2,\,\cal C).\nonumber
\end{eqnarray}
Here the matrix $R$ (the spectral transform) is a distribution in
${\cal C}^2$ and a time dependence is given by the following linear
evolution equation:
$$
\partial_tR(k,\,l)=R(k,\,l)\Omega(k)-\Omega(l)R(k,\,l).
\eqno(2)$$
In the above equation $\Omega(k)$ is a matrix-valued function on $\cal C$
called a dispersion relation. In a general case, $\Omega(k)$ consists of
a holomorphic (polynomial) part $\Omega_p(k)$ and a non-analytic (singular)
part $\Omega_s(k)$, i.e. $\bar\partial\Omega_s\ne 0$ in a some subset of
the plane $\cal C$.

Let us denote the integral in (1) as $\phi(k)R_k\,F$ where $F$ is an integral
operator acting on the left in accordance with the equation (1). Hence, we
write (1) as
$$
\bar\partial\phi(k)=\phi(k)R_k\,F.
\eqno(3)$$
A solution of the $\bar\partial$-problem is given by a solution of the linear
integral equation
$$
\phi(k)=1+{1\over2\pi i}\int\!\int\,{{dl\wedge d\bar l}\over{l-k}}
\int\!\int\,dm\wedge d\bar m\,\phi(m)R(l,\,m)
$$
$$=1+{1\over2\pi i}\int\!\int\,{{dl\wedge d\bar l}\over{l-k}}\left(\phi(l)
R_l\,F\right)\equiv1+\phi(k)R_k\,FC_k.
\eqno(4)$$
The integral operator $C_k$ acting on the left transforms the argument $k$
of the function in front of it to $l$ and integrates with the weight
$(2\pi i)^{-1}(l-k)^{-1}$ on the whole complex plane. The integral operators
introduced in such a way allow us to write formally a solution of the
$\bar\partial$-problem (3) as
$$
\phi(k)=1\cdot(1-R_k\,FC_k)^{-1}.
\eqno(5)$$
The similar representation for solutions of the $\bar\partial$-problem was
effectively used by Beals and Coifman \cite{8} in the case of holomorphic
functions $\Omega_p(k)$.

Let us define a pairing for matrix-valued functions on $\cal C$:
$$
\langle\phi,\,\psi\rangle={1\over2\pi i}\int\!\int\,dk\wedge d\bar k\,\phi(k)\,
\tilde\psi(k),
$$
where tilde stands for transpose. With respect to this pairing we have
$$
\langle\phi\,C_k,\,\psi\rangle=-\langle\phi,\,\psi C_k\rangle,
 \qquad \langle\phi R_k\,F,\,\psi\rangle=
\langle\phi,\,\psi\hat R_k\,F\rangle,
\eqno(6)$$
where $\hat R(k,\,l)=\tilde R(l,\,k)$. Assume then a parametric dependence
of $R(k,\,l)$ on spatial variables $(x,\,y)$ of the form
$$\partial_xR(k,\,l)=il\sigma_3R(k,\,l)-ikR(k,\,l)\sigma_3,
$$
$$
\partial_yR(k,\,l)=i(k-l)R(k,\,l).
\eqno(7)$$
Taking as a basis the representation (5), it is shown in Appendix 1 that
the choice (7) is equivalent to setting the two-dimensional
Zakharov-Shabat spectral problem
$$
T_1\phi\equiv(\partial_x+\sigma_3\partial_y+Q)\phi-ik[\sigma_3,\,\phi]=0,
\eqno(8)$$
where a potential $Q$ is defined as
$$
Q(x,\,y)=-i[\sigma_3,\,\langle\phi R_k\,F\rangle]
\eqno(9)$$
and we denote $\langle f,\,1\rangle\equiv\langle f\rangle$.

Now we turn to an evolution linear problem $\partial_t\phi=W\phi+\phi\,\Omega$.
It follows from (5) and (2) that
$$
\phi_t=\phi\,\partial_tR_k\,FC_k(1-R_k\,FC_k)^{-1}=(\phi R_k\,F\Omega\,C_k-
\phi\,\Omega R_k\,FC_k)(1-R_k\,FC_k)^{-1}$$
$$
=(\phi R_k\,F\Omega\,C_k-\phi\,\Omega)(1-R_k\,FC_k)^{-1}+\phi\,\Omega,
$$
which gives
$$
W\phi=(\phi\,\Omega R_k\,FC_k-\phi\,\Omega)(1-R_k\,FC_k)^{-1}.
\eqno(10)$$
It is shown in Appendix 2 that the formula (10) is reduced for $\Omega_p=0$
to
$$
W(k)\phi(k)=-\phi(k)\bar\partial\,\Omega_s(k)C_k(1-R_k\,FC_k)^{-1}.
$$
Multiplying this relation on the right by $(1-R_k\,FC_k)$ and acting by the
$\bar\partial$-operator, we obtain
$$
\bar\partial W\phi+W(\phi R_k\,F)-W\phi R_k\,F=-\phi\,\bar\partial\,\Omega_s,
$$
which gives the integral equation for the function $W$ \cite{3}:
$$
\bar\partial W(k)=-\phi\,\bar\partial\,\Omega_s\phi^{-1}(k)+
\int\!\int dl\wedge d\bar l\,\left[W(l)-W(k)
\right]\phi(l)R(k,\,l)\phi^{-1}(k).
\eqno(11)$$
Hence, the function $W$ is known only to withih a solution of the integral
equation (11). Nevertheless, Boiti et al have shown \cite{3} that it is
possible to derive SDR equations from the corresponding Lax representation
with the operators $T_1$ and $T_2=\partial_t-W$.

It should be noted here that the equation (11) includes the inverse function
$\phi^{-1}$. However, there is no in 2+1 dimensions, contrary to 1+1, a
simple equation (like (8)) for $\phi^{-1}$. Hence, a problem arises to find
a natural (2+1)-dimensional analogue of the inverse function in 1+1
dimensions. We shall show in the following section that such a function does
exist and it permits to simplify the SDR equations in \cite{3}.
\bigskip

\noindent {\bf3. Hierarchy and recursion operator}
\medskip

\noindent
Let us calculate the evolution of the potential $Q$ which is given explicitly
by (9):
$$
\partial_tQ=-i[\sigma_3,\,\langle\partial_t\,(\phi R_k\,F)\rangle].
$$
The right-hand side can be transformed as follows:
$$
\partial_t\,(\phi R_k\,F)=\partial_t\phi R_k\,F+\phi\,\partial_tR_k\,F=
W\phi R_k\,F+\phi R_k\,F\Omega.
$$
Further calculation, due to (10), yields
\begin{eqnarray*}
\partial_t(\phi R_k\,F)&=&\phi R_k\,F\Omega C_k(1-R_k\,FC_k)^{-1}R_k\,F-
\phi\,\Omega(1-R_k\,FC_k)^{-1}R_k\,F\\
 +\phi R_k\,F\Omega&=&\phi R_k\,F\Omega(1-C_k\,R_k\,F)^{-1}-\phi\,\Omega R_k
\,F(1-C_k\,R_k\,F)^{-1}.
\end{eqnarray*}
Hence,
$$
\partial_tQ=-i[\sigma_3,\,\langle\phi R_k\,F\Omega(1-C_k\,R_k\,F)^{-1},\,1
\rangle-\langle\phi\,\Omega R_k\,F(1-C_k\,R_k\,F)^{-1}\rangle].
$$
Taking into account properties (6) of the pairing, we get
$$
\partial_tQ=-i[\sigma_3,\,\langle\phi R_k\,F\Omega,\,1\cdot(1+\hat R_k\,FC_k
)^{-1}\rangle-\langle\phi\,\Omega,\,1\cdot(1+\hat R_k\,FC_k)^{-1}\hat R_k\,F
\rangle].
\eqno(12)$$
Now we introduce a dual function $\phi^*(k)$ by means of the relation
$$
\tilde\phi^*(k)=1\cdot(1+\hat R_k\,FC_k)^{-1}.
\eqno(13)$$
The $\bar\partial$-problem for the dual function has the form
$$
\bar\partial\phi^*(k)=-\int\!\int dl\wedge d\bar l\,R(l,\,k)\phi^*(l), \qquad
\bar\partial\,\tilde\phi^*(k)=-\tilde\phi^*(k)\hat R_k\,F
\eqno(14)$$
and $\phi^*(k)$ satisfies the dual spectral problem
$$
\partial_x\phi^*+\partial_y\phi^*\sigma_3-\phi^*Q-ik[\sigma_3,\,\phi^*]=0.
\eqno(15)$$
The derivation of (14) and (15) is given in Appendix 3. It is seen from (15)
that just the dual function $\phi^*$ is a true (2+1)-dimensional
generalisation of inverse functions in 1+1 dimensions. It should be stressed
that the definition (13) of the dual function arises naturally within the
framework of the formalism based on the representation (5).

Taking into account the above relations concerning the dual function, we
write the evolution (12) in the form
\begin{eqnarray*}
\partial_tQ&=&-i[\sigma_3,\,\langle\phi R_k\,F\Omega,\,\tilde\phi^*\rangle
-\langle\phi\,\Omega,\,
\tilde\phi^*\hat R_k\,F\rangle]\\
&=&-i[\sigma_3,\,\langle\bar\partial\phi\,\Omega\,\phi^*\rangle
+\langle\phi\,\Omega\,\bar\partial\phi^
*\rangle].
\end{eqnarray*}
Finally, dividing the dispersion relation into regular and singular parts,
we obtain under condition $\Omega_s(k)\rightarrow0$
for $k\rightarrow\infty$:
$$
\partial_t Q=-i\lbrack\sigma_3,\,\langle\bar\partial(\phi\Omega_p\phi^*)
\rangle-\langle\phi\,\bar\partial\,\Omega_s\,\phi^*\rangle].
\eqno(16)$$
Assume further that
$$
\Omega_p(k)=\alpha_nk^n\sigma_3, \qquad \alpha_n=\mbox{const}, \qquad n=0,\,1,
\ldots,$$
$$
\Omega_s(k)={1\over2\pi i}\int\!\int\,{dl\wedge d\bar l\over
l-k}g(l)\sigma_3,
$$
which yields $\bar\partial\,\Omega_s(k)=g(k)\sigma_3$. Now we introduce a
bilocal object
$$
M_{12}(x,\,y_1,\,y_2,\,k)=\phi(x,\,y_1,\,k)\sigma_3\,\phi^*(x,\,y_2,\,k)
\equiv\phi_1\sigma_3\,\phi^*_2
$$
It is easy to see that the function $M_{12}$ satisfies the equation
$$
\partial_xM_{12}+\sigma_3\partial_{y_1}M_{12}+\partial_{y_2}M_{12}\sigma_3
-ik[\sigma_3,\,M_{12}]+Q_1M_{12}-M_{12}Q_2=0,
\eqno(17)$$
where $Q_i\equiv Q(x,\,y_i)$, $i=1,\,2$. Hence, equation (16) takes the form
$$
\delta_{12}\,\partial_tQ_2=
-i\alpha_n\delta_{12}[\sigma_3,\,\langle\bar\partial\,
(k^nM_{12})\rangle]+i\delta_{12}[\sigma_3,\,\langle g(k)M_{12}\rangle],
\eqno(18)$$
where $\delta_{12}=\delta(y_1-y_2)$. Following \cite{10}, we introduce
the notations
$$
P_{12}M_{12}=\partial_xM_{12}+\sigma_3\partial_{y_1}M_{12}+\partial_{y_2}
M_{12}\sigma_3, \quad Q^{\pm}_{12}M_{12}=Q_1M_{12}\pm M_{12}Q_2.
\eqno(19)$$
Let $M_{12}^d$ and $M_{12}^a$ are the diagonal and off-diagonal parts of the
matrix $M_{12}$, respectively. Then (17) and (19) yield
$$
P_{12}\,M_{12}^d+Q_{12}^-\,M_{12}^a=0,
\eqno(20)$$
$$
P_{12}\,M_{12}^a-2\,ik\sigma_3\,M_{12}^a+Q_{12}\,M_{12}^d=0.
\eqno(21)$$
We can write from (20) the diagonal part as $M_{12}^d=\sigma_3-P_{12}^
{-1}Q_{12}^-M_{12}^a$. Hence, the equation (21) is written in the form
$(\Lambda-k)M_{12}^a=(2i)^{-1}Q_{12}^+\cdot1$, where the
operator $\Lambda$ is defined as
$$
\Lambda={1\over2i}\sigma_3(P_{12}-Q_{12}^-\,P_{12}^{-1}Q_{12}^-).
$$
Then $M_{12}^a=(2i)^{-1}(\Lambda-k)^{-1}\,Q_{12}^+\cdot1$
and after expansion $(\Lambda-k)^{-1}
=-\sum_{m=1}^\infty k^{-m}\Lambda^{m-1}$
we can write the polynomial contribution to $\partial_tQ$ in (18) as
$$
-i\alpha_n\delta_{12}[\sigma_3,\,\langle\bar\partial\,(k^nM_{12})\rangle]=
\alpha_n\sigma_3\,\delta_{12}\sum\limits_{m=1}^\infty\langle\bar\partial
k^{n-m}\rangle\Lambda^{m-1}Q_{12}^+\cdot1
$$
$$=-{i\over2}\alpha_n\sigma_3\,\delta_{12}\,\Lambda^nQ_{12}^+\cdot1.
$$
Now we have all the needed to formulate a closed system of equations
describing the evolution of the potential $Q$ under condition of the linear
evolution of the spectral transform $R$:
$$
\delta_{12}\,\partial_tQ_2=
-{i\over2}\alpha_n\sigma_3\,\delta_{12}\,\Lambda^nQ_{12}^+
\cdot1+i\delta_{12}[\sigma_3,\,\langle g(k)M_{12}\rangle],
$$
$$
\left(P_{12}\,M_{12}-ik[\sigma_3,\,M_{12}]+Q_{12}^-\,M_{12}\right)g(k)=0.
\eqno(22)$$
Here the operator $\Lambda$ plays the role of a recursion operator (more
precisely, $\Lambda$ is connected with the true recursion operator by means
of $\sigma_3$ \cite{10}). If $g(k)=0$, we get from
(22) the well-known hierarchy including the Davey-Stewartson-1 equation
derived by Santini and Fokas \cite{10} on the basis of an integral
representation for $W$.

For $\Omega_p=0$ the system (22) takes the form
$$
\delta_{12}\,\partial_tQ_2=i\delta_{12}[\sigma_3,\,\langle g(k)M_{12}\rangle],
$$
$$
\left(P_{12}\,M_{12}-ik[\sigma_3,\,M_{12}]+Q_{12}^-\,M_{12}\right)g(k)=0.
\eqno(23)$$
It is seen that the structure of the system (23) is similar to that for
(1+1)-dimensional Maxwell-Bloch equations $(g(k)\sim\delta(\Im k)
\delta(\Re k-\alpha))$:
$$
\partial_tQ=i[\sigma_3,\,\langle g(k)\Phi(k)\rangle], \qquad \Phi=\phi\,
\sigma_3\,\phi^{-1},
$$
$$
\partial_x\Phi(\alpha)-i\alpha[\sigma_3,\,\Phi(\alpha)]+[Q,\,\Phi(\alpha)]=0
\eqno(24)$$
and the system (23) is reduced to (24) in the (1+1)-dimensional limit.
It should be stressed that, as distinct from \cite{3}, the (2+1)-dimensional
counterpart (23) of the Maxwell-Bloch equations demonstrates explicitly the
presence of the 'squared function' term. It can be shown that a function
$\Gamma$ introduced in \cite{3} is expressed, as a matter of fact, in terms
of the above squared functions as $\Gamma=i\bar\partial\,(k\phi\,\phi^*)$.
Finally, for $n=2$ equations (22) yield a (2+1)-dimensional generalisation
of the equations derived in \cite{12} in the context of non-linear optics.
\bigskip

\noindent {\bf4. Conclusion}
\medskip

\noindent
We proposed a procedure for obtaining (2+1)-dimensional non-linear equations
with non-analytic dispersion relations compatible with the linear evolution
of the spectral transform. An important step in deriving these equations was
to use the representation (5) of the $\bar\partial$-problem solution. In
spite of the formality of this representation, it allows us to perform all
the needed manipulations. The introduction of the dual functions has made it
possible to obtain the hierarchy of equations without explicit use of the
second Lax operator. The application of the bilocal formalism was crucial
for bringing these equations to a form similar to that for their
counterparts in 1+1 dimensions.
\bigskip

\noindent {\bf Acknowledgment}
\medskip

\noindent
The author thanks Dr V G Dubrovsky for helpful discussions.
\newpage

\noindent {\bf Appendix 1. Linear spectral problem}
\medskip

\noindent
We show here that the choice (7) of the dependence of $R(k)$ on spatial
variables $x$ and $y$ leads to the Zakharov-Shabat problem on the plane.
Differentiating (5) with respect to $x$ we obtain $\partial_x\phi=
\phi\,\partial_xR_k\,FC_k(1-R_k\,FC_k)^{-1}$.
In virtue of the definitions of
the integral operators $F$ and $C_k$ and (7) we can perform the following
calculation:
$$
\phi\,\partial_xR_k\,FC_k={1\over2\pi i}\int\!\int{dl\wedge d\bar l\over l-k}
\int\!\int dm\wedge d\bar m\,\phi(m)\partial_xR(l,\,m)
$$
$$
={1\over2\pi i}\int\!\int{dl\wedge d\bar l\over l-k}\int\!\int dm\wedge d\bar
m\,im\,\phi(m)\sigma_3\,R(l,\,m)
$$
$$
-{1\over2\pi i}\int\!\int {dl\wedge d\bar l
\over l-k}\,il\,\int\!\int dm\wedge d\bar m\,\phi(m)\,R(l,\,m)\sigma_3
$$
$$
={1\over2\pi i}\int\!\int{dl\wedge d\bar l\over l-k}(il\,\phi\sigma_3\,
R_l\,F)-{1\over2\pi i}\int\!\int dl\wedge d\bar l\,i\left(1+{k\over l-k}
\right)\,(\phi\,R_l\,F)\sigma_3.
\eqno(\mbox{A}.1)$$
Since we have from (4) $\phi R_k\,FC_k=\phi-1$, then (A.1) and evident
relation $R_k\,FC_k(1-R_k\,FC_k)^{-1}=(1-R_k\,FC_k)^{-1}-1$ yield
$$
\partial_x\phi=-ik\,\phi\sigma_3-i\langle\phi\,R_k\,F\rangle\sigma_3\,\phi+
ik\,\sigma_3(1-R_k\,FC_k)^{-1}.
\eqno(\mbox{A}.2)$$
Similarly,
$$
\partial_y\phi=ik\,\phi+i\langle\phi R_k\,F\rangle\phi-ik\,(1-R_k\,FC_k)^{-1}.
\eqno(\mbox{A}.3)$$
Adding (A.2) and (A.3) yields
$$
\partial_x\phi+\sigma_3\,\partial_y\phi-ik\,[\sigma_3,\,\phi]-i[\sigma_3,\,
\langle\phi R_k\,F\rangle]\phi=0.
\eqno(\mbox{A}.4)$$
Hence, if we identify $-i[\sigma_3,\,\langle\phi R_k\,F\rangle]\equiv Q(x,\,
y)$, (A.4) gives the above spectral problem.
\bigskip

\noindent {\bf Appendix 2. Linear evolution problem}
\medskip

\noindent
In order to derive the linear  evolution problem $\partial_t\phi=W\phi+\phi\,
\Omega$, we calculate $\partial_t\phi$ from (2) and (5). Let us take for
simplicity $\Omega_p=0$ whereas
$$
\Omega(k)=\Omega_s(k)={1\over2\pi i}\int\!\int{ds\wedge d\bar s\over s-k}
g(s)\sigma_3,
$$
which gives $\bar\partial\,\Omega(k)=g(k)\sigma_3$. The calculation yields
$$
\partial_t\phi=\phi\,\partial_tR_k\,FC_k(1-R_k\,FC_k)^{-1}$$
$$
=\lbrack\phi R_k\,F\,\Omega\,C_k-\phi\,\Omega\,R_k\,FC_k\rbrack(1-R_k\,FC_k)
^{-1}$$
$$
=\phi R_k\,F\,\Omega\,C_k(1-R_k\,FC_k)^{-1}-\phi\,\Omega(1-R_k\,FC_k)^{-1}
+\phi\,\Omega$$
$$
\equiv W(k)\phi(k)+\phi(k)\,\Omega(k),
$$
where
$$
W(k)\phi(k)=(\phi R_k\,F\,\Omega\,C_k-\phi\,\Omega)(1-R_k\,FC_k)^{-1}.
\eqno(\mbox{B}.1)$$
Taking into account the definitions of the integral operators $F$ and $C_k$,
we can rewrite (B.1) as
$$
W(k)\phi(k)(1-R_k\,FC_k)={1\over2\pi i}\int\!\int{dl\wedge d\bar l\over l-k}
\left(\phi(l)R_l\,F\right)\Omega(l)-\phi\,\Omega
$$
$$
={1\over2\pi i}\int\!\int{dl\wedge d\bar l\over l-k}\int\!\int dm\wedge d\bar m
\,\phi(m)R(l,\,m){1\over2\pi i}\int\!\int{ds\wedge d\bar s\over s-l}g(s)
\sigma_3-\phi\,\Omega.
\eqno(\mbox{B}.2)$$
The denominator in (B.2) can be represented as
$$
{1\over(l-k)(s-l)}={1\over s-k}\left({1\over l-k}-{1\over l-s}\right).
$$
Then we have $$W\phi\,(1-R_k\,FC_k)$$
$$
={1\over2\pi i}\int\!\int{dl\wedge d\bar l\over l-k}
\int\!\int dm\wedge d\bar m\,\phi(m)R(l,\,m){1\over2\pi i}\int\!\int
{ds\wedge d\bar s\over s-k}g(s)\sigma_3
$$
$$
-{1\over2\pi i}\int\!\int{ds\wedge d\bar s\over s-k}g(s){1\over2\pi i}
\int\!\int{dl\wedge d\bar l\over l-s}\int\!\int dm\wedge d\bar m\,\phi(m)
R(l,\,m)\sigma_3-\phi\,\Omega$$
$$
=\phi R_k\,FC_k\,\Omega(k)-{1\over2\pi i}\int\!\int{ds\wedge d\bar s\over s-k}
g(s)(\phi R_s\,FC_s)\sigma_3-\phi\,\Omega$$
$$
=-g(k)\phi(k)\sigma_3\,C_k,
$$
where, as in Appendix 1, we use $\phi R_s\,FC_s=\phi(s)-1$. Hence,
$$
W(k)\phi(k)=-g(k)\phi(k)\sigma_3\,C_k(1-R_k\,FC_k)^{-1}.
$$

It should be noted that the calculation of $\partial_t\phi$ for $\Omega_s=0$
and $\Omega_p=\alpha_2k^2\sigma_3$ on the basis of (5) leads to the
well-known operator $T_2=\partial_t-W$ for the Davey-Stewartson-1 equation
\cite {5,12}, where a potential $A$ of the mean flow has the form
$$
A=2\sigma_3\bigl[\langle k(\phi R_k\,F)\rangle-\langle k\phi R_k\,F\rangle
-i\langle\partial_y\phi R_k\,F\rangle\bigr]^d.
$$
\bigskip

\noindent {\bf Appendix 3. Dual spectral problem}
\medskip

\noindent
Here we give a derivation of the equations (14) and (15). The definition (13)
gives $\tilde\phi^*=1-\tilde\phi^*\hat R_k\,FC_k$ and taking into account the
evident property $\bar\partial f(k)C_k=f(k)$ for any function $f(k)$, this
yields
$$
\bar\partial\tilde\phi^*=-\tilde\phi^*\hat R_k\,F=-\int\!\int dl\wedge d\bar
l\,\tilde\phi^*(l)\hat R(k,\,l)$$
$$
=-\int\!\int dl\wedge d\bar l\,\tilde\phi^*(l)\tilde R(l,\,k)=-
\biggl[\int\!\int dl\wedge d\bar l\,R(l,\,k)\phi^*(l)\biggr]_
{\mbox{(transpose)}}
$$
Hence, equation (14) follows. Now we find a spectral problem for the dual
function $\phi^*$. Differentiating (13) with respect to $x$, we find
$\partial_x\tilde\phi^*=-\tilde\phi^*\partial_x\hat R_k\,FC_k
(1+\hat R_k\,FC_k)^{-1}$. Taking into account that $\hat R(k,\,l)=
\tilde R(l,\,k)$, we obtain from (7)
$$
\partial_x\hat R(k,\,l)=ik\,\hat R(k,\,l)\sigma_3-il\,\sigma_3\hat R(k,\,l),
\quad \partial_y\hat R(k,\,l)=-i(k-l)\hat R(k,\,l).
$$
Then  following the calculation in Appendix 1, we obtain
$$
\partial_x\tilde\phi^*=i\tilde\phi^*\sigma_3-i\langle\tilde\phi^*\hat R_k\,
F,\,1\rangle\sigma_3\tilde\phi^*-ik\,\sigma_3(1+\hat R_k\,FC_k)^{-1},
$$
and
$$
\partial_x\phi^*=ik\,\sigma_3\phi^*-i\phi^*\sigma_3\langle1,\,\tilde\phi^*
\hat R_k\,F\rangle-ik\,(1+\hat R_k\,FC_k)^{-1}_{\mbox{(transp)}}\sigma_3.
$$
Similarly,
$$
\partial_y\phi^*=-ik\,\phi^*+i\phi^*\langle1,\,\tilde\phi^*\hat R_k\,F\rangle
+ik\,(1+\hat R_k\,FC_k)^{-1}_{\mbox{(transp)}}.
$$
Hence,
$$
\partial_x\phi^*+\partial_y\phi^*\sigma_3-ik\,[\sigma_3,\,\phi^*]+i\phi^*
[\sigma_3,\,\langle1,\,\tilde\phi^*\hat R_k\,F\rangle]=0.
\eqno(C.1)$$
We need now a connection of $\langle1,\,\tilde\phi^*\hat R_k\,F\rangle$
with $Q$. It can be found as follows:
$$
\langle\phi R_k\,F,\,1\rangle=\langle1\cdot(1-R_k\,FC_k)^{-1}R_k\,F,\,1\rangle
=\langle1\cdot(1-R_k\,FC_k)^{-1},\,\hat R_k\,F\rangle$$
$$
=\langle1,\,\hat R_k\,F(1+C_k\hat R_k\,F)^{-1}\rangle
=\langle1,\,1\cdot(1+\hat R_k
\,FC_k)^{-1}\hat R_k\,F\rangle=\langle1,\,\tilde\phi^*\hat R_k\,F\rangle.
$$
Hence,
$$
Q=-i[\sigma_3,\,\langle\phi R_k\,F,\,1\rangle]=-i[\sigma_3,\,\langle1,\,
\tilde\phi^*\hat R_k\,F\rangle],
$$
and we derive from (C.1) the equation (15).

\end{document}